\documentclass[sigconf]{acmart}
\usepackage{enumitem}
\usepackage{booktabs}
\usepackage{multirow}
\usepackage{amssymb}
\usepackage{mdframed}
\usepackage{pgfplots}
\usepackage{rotating}
 \usepackage{array,graphicx}
 \usepackage{float}
 \usepackage[latin1]{inputenc}
\usepackage[T1]{fontenc}
 
\usepackage{hyperref}

\newenvironment{xquote}{%
  \begin{list}{}{%
    \small%
    \setlength{\leftmargin}{0.2cm}%
    \setlength{\rightmargin}{0pt}%
  }%
  \item\relax
}{\end{list}}

\usepackage[font={small}]{caption}

\newcommand*\circled[1]{\tikz[baseline=(char.base)]{
            \node[shape=circle,fill,inner sep=0.5pt] (char) {\textcolor{white}{#1}};}}
            
\copyrightyear{2020}
\acmYear{2020}
\setcopyright{acmcopyright}\acmConference[ESEM '20]{ESEM '20: ACM / IEEE International Symposium on Empirical Software Engineering and Measurement (ESEM)}{October 8--9, 2020}{Bari, Italy}
\acmBooktitle{ESEM '20: ACM / IEEE International Symposium on Empirical Software Engineering and Measurement (ESEM) (ESEM '20), October 8--9, 2020, Bari, Italy}
\acmPrice{15.00}
\acmDOI{10.1145/3382494.3421462}
\acmISBN{978-1-4503-7580-1/20/10}            

\begin{document}

\title{What Makes Agile Test Artifacts Useful? An Activity-Based Quality Model from a Practitioners' Perspective%
}

\author{Jannik Fischbach}
\author{Henning Femmer}
\affiliation{%
\institution{Qualicen GmbH}
}
\email{firstname.lastname@qualicen.de}
\author{Daniel Mendez}
\additionalaffiliation{%
  \institution{fortiss GmbH}
}
\author{Davide Fucci}
\affiliation{%
\institution{Blekinge Institute of Technology}
}
\email{firstname.lastname@bth.se}
\author{Andreas Vogelsang}
\affiliation{%
\institution{University of Cologne}
}
\email{vogelsang@cs.uni-koeln.de}

\begin{abstract}
    \textbf{Background:} The artifacts used in Agile software testing and the reasons why these artifacts are used are fairly well-understood.
    However, empirical research on how Agile test artifacts are eventually designed in practice and which quality factors make them useful for software testing remains sparse.
    \textbf{Aims:} Our objective is two-fold. First, we identify current challenges in using test artifacts to understand why certain quality factors are considered good or bad. Second, we build an Activity-Based Artifact Quality Model that describes what Agile test artifacts should look like.
    \textbf{Method:} We conduct an industrial survey with 18 practitioners from 12 companies operating in seven different domains. 
    \textbf{Results:} Our analysis reveals nine challenges and 16 factors describing the quality of six test artifacts from the perspective of Agile testers. Interestingly, we observed mostly challenges regarding language and traceability, which are well-known to occur in non-Agile projects. 
    \textbf{Conclusions:} Although Agile software testing is becoming the norm, we still have little confidence about general \textit{do's} and \textit{don'ts} going beyond conventional wisdom.
    This study is the first to distill a list of quality factors deemed important to what can be considered as useful test artifacts.
\end{abstract}



\maketitle 

\section{Introduction}
The Agile Software Development (ASD) principle ``working software over comprehensive documentation'' promotes that documentation should be kept to what is necessary or useful~\cite{beck2001}. 
Hence, common ASD frameworks, such as Scrum~\cite{Schwaber95}, mention only few artifacts (\textsc{Epics}, \textsc{User story}, etc.) that should be created, used, and maintained for documentation purposes. 
Instead, face-to-face communication should be encouraged in order to convey information.
Nevertheless, Agile practitioners have increasingly changed their attitude towards documentation ~\cite{Stettina11} and are producing a variety of artifacts that are not inherent to ASD ~\cite{Bass16,liskin15,wagenaar15}. 
According to Wagenaar et al.~\cite{Wagenaar2018}, practitioners need additional artifacts for four reasons: i) they provide team governance, ii) they are useful for internal communication, iii) they are needed by external parties, and iv) they are useful for quality assurance.
For the latter reason, a range of additional artifacts (e.g., \textsc{Acceptance tests}) are commonly created to perform comprehensive software testing. 

Currently, we understand \textit{which} test artifacts Agile teams introduce (or should introduce) on their own initiative~\cite{Bass16} and \textit{why} they are needed~\cite{Wagenaar2018}. 
However, empirical research on \textit{how} Agile test artifacts are designed in practice and, more specifically, \textit{which} properties make them useful for quality assurance remains sparse.
Existing normative standards such as the ISTQB Acceptance Testing Syllabus~\cite{ISTQB} or ISO 29119:2013~\cite{ISO} occasionally mention some properties that test artifacts should possess.
However, there are issues with these normative standards.
Firstly, the list of properties is not complete---most of the properties are defined for the artifacts introduced by the ASD frameworks but not for the additionally-required artifacts introduced by the team.
Secondly, normative standards describe quality through abstract properties---e.g., \textsc{Acceptance criteria} should be both ``precise and concise''~\cite{ISTQB}.
The standard does not provide any further description of what is meant by these vague properties. 
Thirdly, the empirical basis and reasoning for these criteria remains unclear.
This implies that the criteria are difficult, if not impossible to falsify. We argue that for a combination of all of these reasons, we observe in practice that Agile teams fail to satisfy these normative criteria, and struggle in maintaining their documentation artefacts~\cite{hotomski}. 

In contrast to these existing ways to define quality criteria, we argue that quality of test artifacts should be defined from a quality-in-use perspective. 
Following the idea of Activity-Based Artifact Quality Models~\cite{qualityInUse19}, we postulate that the quality of a test artifact depends on the stakeholder using it and the activities for which it is used. 
Accordingly, we \textbf{explore properties (so-called ``quality factors'') of test artifacts that have a positive or negative impact on the activities of the stakeholders}. 
To understand why a certain quality factor is considered good or bad by the practitioners, we first study current challenges in using test artifacts. 
Consequently, we extend the normative qualities with a list of concrete factors describing what Agile test artifacts should look like.
For this purpose, we conduct an industrial survey based on one-on-one interviews with 18 practitioners from 12 companies operating in seven domains, and make the following contributions (C):
\begin{enumerate}[label=\bfseries C \arabic*:,leftmargin=*]
    \item A list of nine challenges that practitioners face when using the test artifacts \textsc{Acceptance criterion}, \textsc{Acceptance test}, \textsc{Feature}, \textsc{Test documentation}, \textsc{Test data}, and \textsc{Unit tests}.
    \item An Activity-Based Quality Model of 16 quality factors for these artifacts, serving as a foundation for systematic quality control in practice.
\end{enumerate}




\section{Fundamentals}
\begin{figure}
    \begin{mdframed}
        \includegraphics[width=\textwidth]{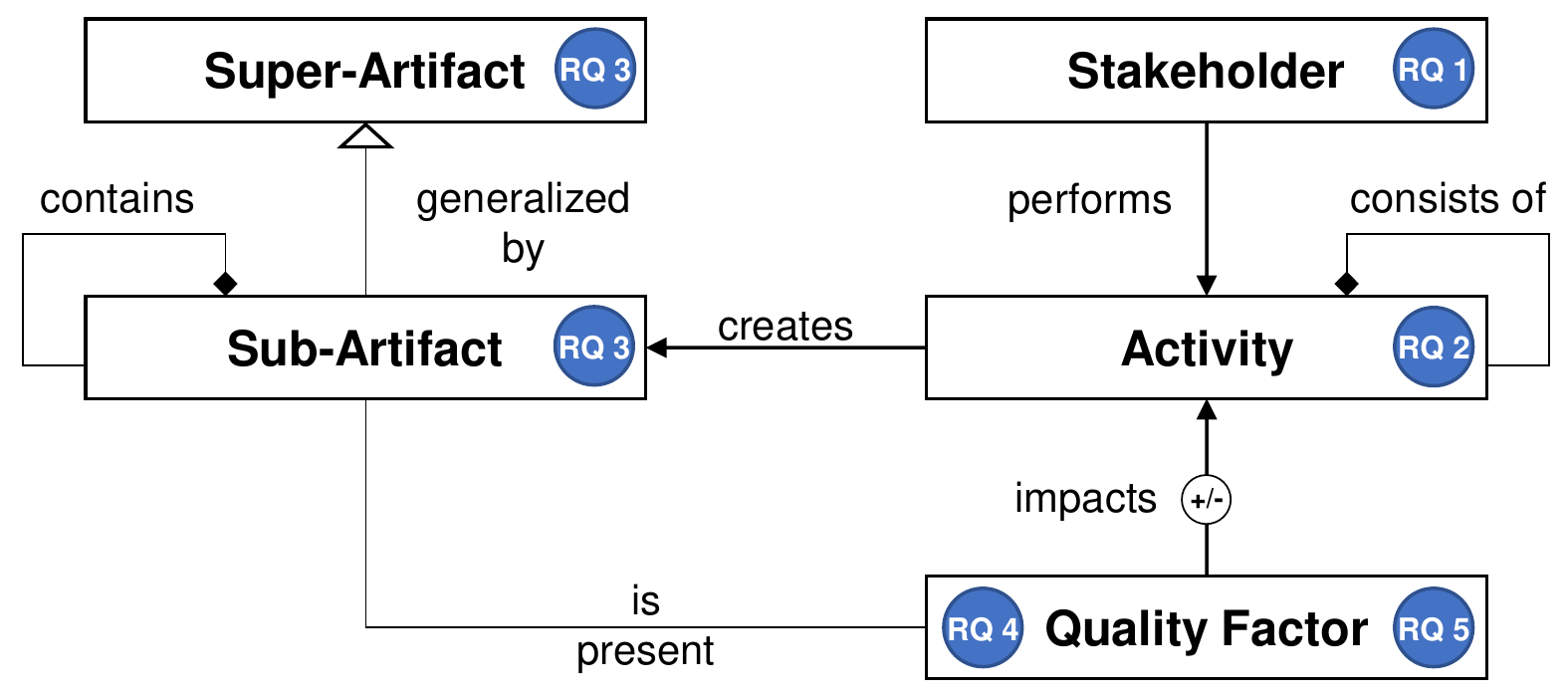}
    \end{mdframed}
    \vspace{-.2cm}
    \caption{ABAQM meta model and mapped RQs, based on~\cite{femmer2017rapid}}
    \label{fig:metamodel}
    \vspace{-.5cm}
\end{figure}

In this section, we briefly define the theory that we will use as the foundation of this work. Femmer and Vogelsang~\cite{qualityInUse19} argue that it is not sufficient to speak of \textit{good and bad quality} in general since the quality of an artifact depends on the context in which it is used. 
More specifically, quality is determined by the stakeholders and the activities that they conduct with the artifact. The quality of an artifact is considered good if its properties allow stakeholders to effectively and efficiently carry out their activities. 
Following this line of thought, they propose Activity-based Artifact Quality Models (ABAQM) and apply them to study the quality of requirements engineering artifacts~\cite{femmer2017rapid}.
In this paper, we create an ABAQM for all test artifacts involved in ASD, which enables us to understand their quality in the Agile context. 
We use the following concepts to describe an ABAQM (see a meta model in Fig.~\ref{fig:metamodel}):

\textbf{Artifact:} Following the quality-in-use paradigm, an artifact is a collection of coherent documented information which assists a stakeholder in reaching the project goals.
Examples of artifacts are \textsc{Use case documents} and \textsc{Test data}. 
Artifacts that share similar properties can be combined into a generalized super-class. 
For example, \textsc{Unit tests}, \textsc{Integration tests} and \textsc{System tests} address different test levels but can be bundled into a super-class \textsc{Test}. 
In addition, artifacts can contain other artifacts, such as a \textsc{User story} which contains multiple \textsc{Acceptance criteria}.

\textbf{Stakeholder:} A stakeholder is interested in an artifact and uses it during a certain activity. 
An example of a stakeholder is a \textsc{Test designer}, who uses \textsc{User stories} to derive \textsc{Acceptance tests}. 

\textbf{Activity:} An activity is an invested effort which involves one or more of the mentioned artifacts. 
An activity can be divided into sub-activities.
For example, \textsc{Acceptance test design} can be decomposed into \textsc{Acceptance test creation} and \textsc{Acceptance test updating}. 
During an activity, stakeholders do not only use artifacts but also create new ones.
Hence, artifacts can be both input and output of activities.

\textbf{Quality Factor:} A quality factor is a property that is or is not present in an artifact. 
Femmer and Vogelsang stress that this property ``\textit{must be objectively assessable through a measure to be used for quality control}''~\cite{qualityInUse19}. 
For example, a \textsc{Test} should only contain the minimum number of required \textsc{Test cases} to avoid excessive testing.
This quality factor \textsc{Minimal} can be evaluated objectively.

\textbf{Impact:} An impact is a relation between a quality factor and an activity. 
The relation can be either positive (i.e., the presence of the quality factor supports the stakeholder in the execution of an activity) or negative (i.e. the quality factor hinders the stakeholder). 
The aforementioned quality factor \textsc{Minimal}, for example, has a positive impact on the activity \textsc{Testing}.

\section{Methodology}
In order to identify and understand the quality factors of Agile test artifacts, we chose (qualitative) survey as our research method. 
For our study, we followed the guidelines by Ciolkowski et al.~\cite{Ciolkowski2003} for conducting empirical studies based on surveys. 
These guidelines include six steps that are performed in an iterative fashion: \textit{definition}, \textit{design}, \textit{implementation}, \textit{execution}, \textit{analysis}, and \textit{packaging}.

\subsection{Survey Definition}

\subsubsection{Goal of this study}
Following the Goal-Question-Metric~\cite{GQM} technique, we define the goal of our survey as follows:

\begin{enumerate}
    \item \textit{Object}. Test artifacts
    \item \textit{Purpose}. Identify, understand, and define
    \item \textit{Focus}. Quality factors
    \item \textit{Viewpoint}. Agile practitioners
    \item \textit{Context}. Agile Software Development Projects
\end{enumerate}

The expected outcome of our survey is a better understanding of quality factors of test artifacts. 
In our activity-based quality understanding, these are properties that positively or negatively affect the stakeholders and their follow-up activities. 
These quality factors should provide guidance for practitioners on how test artifacts should be designed. 
It should further establish the foundation for a systematic quality control of test artifacts.
Based on the classification of Robson~\cite{Robson2002}, our research goal is exploratory as we are seeking for new insights into the quality of Agile test artifacts.

\subsubsection{Research Questions}
Based on the idea that artifact quality is determined by the context in which it is used, we derived five research questions (RQ) from our survey goal.
Each RQ addresses a specific component of the ABAQM meta model (see Fig.~\ref{fig:metamodel}).

\begin{enumerate}[label=\bfseries RQ \arabic*:,leftmargin=*]
    \item Which stakeholders are involved in Agile testing?
    \item Which activities are performed by the stakeholders?
    \item Which artifacts are used by the stakeholders in the context of these activities?
    \item Which quality factors positively influence the execution of these activities?
    \item Which quality factors negatively influence the execution of these activities?
\end{enumerate}

\subsection{Survey Design}

\subsubsection{Population and survey sample}
The selection of the survey participants was driven by a purposeful sampling strategy~\cite{patton90}.
Specifically, we defined criteria that the participants need to meet to be suitable for our survey, a) they work for a company that develops software following a defined Agile software paradigm (e.g., SCRUM, SAFe), b) they have been involved in the testing process for at least one year, and c) they create, use and/or maintain at least one test artifact. 
Each researcher involved in the survey prepared a list of potential interview partners using their industrial contacts (convenience sampling).
From this list, the research team jointly selected suitable partners based on their adequacy for the study. 
To further increase the sample size, we asked each interviewee for relevant contacts after the interview (snowball sampling). 
We stopped conducting more interviews after we reached saturation. More precisely, once we could no longer identify new quality factors.
Tab.~\ref{tab:participants} presents an overview of the participants, their roles, and information about their companies.
In total, 18 practitioners from 12 different companies operating in seven different domains participated in our survey. 
We did not restrict our population with regard to company size or application domain.
Rather, we involved practitioners from companies of different domains and sizes to obtain a holistic understanding of test artifact quality.

\begin{table}[]\caption{List of Participants}
\label{tab:participants}
\renewcommand{\arraystretch}{0.8}
    \vspace{-.3cm}
\begin{tabular}{@{}lllll@{}}
\toprule
Company             & No. & Role          & Size                               & Domain                     \\ \midrule
\multirow{2}{*}{C1} & P1          & Product Owner & \multirow{2}{*}{150k} & \multirow{2}{*}{Insurance} \\
                    & P2          & Test Designer &                                    &                            \\
\multirow{2}{*}{C2} & P3          & Test Designer & \multirow{2}{*}{1k} & \multirow{2}{*}{Retail} \\
                    & P4          & Test Lead &                                    &                            \\ 
            C3      & P5          & Test Lead &   20             &        Software                    \\ 
            C4      & P6          & Agile Team Lead &   50             &        E-Mobility            
            \\ 
           \multirow{3}{*}{C5} & P7          & Partner & \multirow{3}{*}{500} & \multirow{3}{*}{IT Consulting} \\
                    & P8          & Software Developer &                                    &                            \\
                                        & P15          & Software Developer &                                    &                            \\
            C6      & P9          &  Product Owner  &   10             &        PropTech
             \\ 
            C7      & P10          &  Agile Coach  &   50             &        IT Consulting
            \\
             C8      & P11          &  Agile Team Lead  &   1k             &        Software
             \\ 
              C9      & P12          &  Software Developer &   2k             &        Software
             \\ 
              C10      & P13          &  Software Architect &   200             &        Software
               \\ 
             \multirow{2}{*}{C11} & P14          & Software Architect & \multirow{2}{*}{200} & \multirow{2}{*}{Software} \\
                    & P16          & Software Architect &                                    &                            \\ 
                    \multirow{2}{*}{C12} & P17          & Business Analyst & \multirow{2}{*}{40k} & \multirow{2}{*}{Reinsurance} \\
                    & P18          & Business Analyst &                                    &                            \\ 
            \bottomrule
\end{tabular}
\end{table}

\subsubsection{Data Collection}
We chose interviews over other data collection instruments for two reasons.
First, ambiguities in the questions can be resolved directly ensuring that all questions are understood correctly and that they are not skipped.
Second, the interviewer can observe the behavior of the participants and ask them to elaborate their responses (e.g., to better understand the reasoning of the participant, or to go deeper into details). 
This is particularly important to understand why the participant considers the quality of a certain test artifact good or bad.

\subsubsection{Questionnaire Design}\label{questioning} 
Prior to conducting the interviews, we developed an interview guideline to gather the data for answering our RQs.
We designed the interview questions to systematically identify the elements of the ABAQM to shed light on the quality of test artifacts. 
For this purpose, we followed the guidelines of Dillman et al.~\cite{Dillman14} to reduce common mistakes when setting up a questionnaire (e.g., avoiding double-barreled questions). 
Since our research goal and RQ are of exploratory nature, most questions are open-ended.
Our questionnaire consists of 15 questions, including 13 open-ended questions and two closed questions (see Tab.~\ref{tab:questions}). 
In each interview, we asked introductory questions to gather information about the participant's background (e.g., company, experience in ASD), followed by questions about the activities of the participants and the artifacts they use in the context of these activities. 
To avoid misinterpretations, we gave a short briefing of the concepts central to this study at the beginning of each interview. 

The greatest challenge in compiling the questionnaire was to develop questions for determining the quality factors. 
For this purpose, we discussed two different questioning strategies.
First, ask the participant directly which quality factor are important for the artifact to be useful (e.g., ``Which properties should the artifact possess from your perspective?''). 
In this case, the participant is explicitly asked to state the quality factors. 
Second, initially ask the participants which problems occur during their activities and the usage of their artifacts.
Subsequently, use probing questions to ask what exactly bothers the stakeholder about the artifact and how the artifact should have been designed instead. 
Using this ``problem-oriented questioning approach,'' we first determined the problems related to artefact usage and then derived the respective quality factors from these problems.

 \begin{table}[]\caption{Questionnaire Structure}\label{tab:questions}
 \centering
 \vspace{-.3cm}
 \resizebox{\columnwidth}{!}{%
 \begin{tabular}{cl}
 \toprule
 \parbox[t]{2mm}{\multirow{4}{*}{\rotatebox[origin=c]{90}{\textbf{intro}}}} & How many employees work at your company? \\
 & In which domain does the company operate? \\
 & Since when does your project follow the Agile software paradigm? \\
 & Which framework (Crystal, SCRUM etc.) do you follow? \\
 \hline
 \parbox[t]{2mm}{\multirow{7}{*}{\rotatebox[origin=c]{90}{\textbf{core}}}} & What is your role in the testing process? \\
 & Which activities do you perform? \\
 & What is the purpose of your activities? \\
 & Which artifacts do you create as part of your work? \\
 & Which artifacts do you use as part of your work? \\
 & Which artifacts do you maintain as part of your work? \\
 & What do you need the artifacts for? \\
 & Do problems or challenges arise during your activities? \\
 \hline
 \parbox[t]{2mm}{\multirow{3}{*}{\rotatebox[origin=c]{90}{\textbf{probing}}}} & What exactly about the artifact bothers you? \\
 & How should the artifact be designed instead? \\
 & How is the quality of the test artifacts currently checked? \\
 \bottomrule
 \end{tabular}}
 \vspace{-.4cm}
 \end{table}

\subsubsection{Pilot.}
To decide which questioning strategy is best suited for the creation of a ABAQM, we designed a questionnaire for both strategies and evaluated them in a pilot (see step \circled{1} in Fig.~\ref{fig:approach}). 
We conducted the pilot phase iteratively; it consisted of two parts, the internal pilot and the real case pilot.
In the internal pilot, the questions were continuously refined by the research team with regard to suitability, understandability, and correctness. 
The real case pilot involved two interviews with participants from the targeted population, and revealed that the ``problem oriented'' approach is more suitable for collecting the quality factors.
Practitioners struggle to abstract and define independently which property of an artefact leads to good or bad quality.
It proved more effective to gather the quality factors together by first discussing current challenges and then successively deriving the quality factors of the artifacts.
Hence, the probing questions are an integral part of our questionnaire as they encourage the participants to expand a particular anecdote and to define precisely what they like or dislike about the artifact. 

\subsection{Survey Implementation}
During this step, we compiled all the material needed to conduct the survey.
We prepared an invitation letter to ask potential participants for an interview.
In addition, we provided our questionnaire to the participants in advance to allow them to get a first impression of the content of the interview and prepare accordingly.

\subsection{Survey Execution}
All interviews were conducted by the first author.
The duration of the interviews had an average of 41 minutes with a minimum of 31 and a maximum of 67 minutes.
They took place from March to May 2020.
All interviews were conducted remotely via GoToMeeting and Google Hangouts as face-to-face interviews were infeasible due to COVID-19.
We interviewed all participants individually to prevent that their statements were influenced by others.
The participants were informed, before starting the interview, that the data will be treated anonymously.
Additionally, all interviews were conducted in the native language of the participant---German, with the exception of one in English. 
The audio of all interviews was recorded with the permission of the participants for subsequent analysis (see step \circled{2} in Fig.~\ref{fig:approach}). 
Due to confidentiality agreements with the respective individuals, the recordings cannot be published. 

\subsection{Survey Analysis}
\begin{figure*}
    \begin{mdframed}
        \includegraphics[width=\textwidth,trim={0.7cm 0.3cm 0.7cm 0.3cm}]{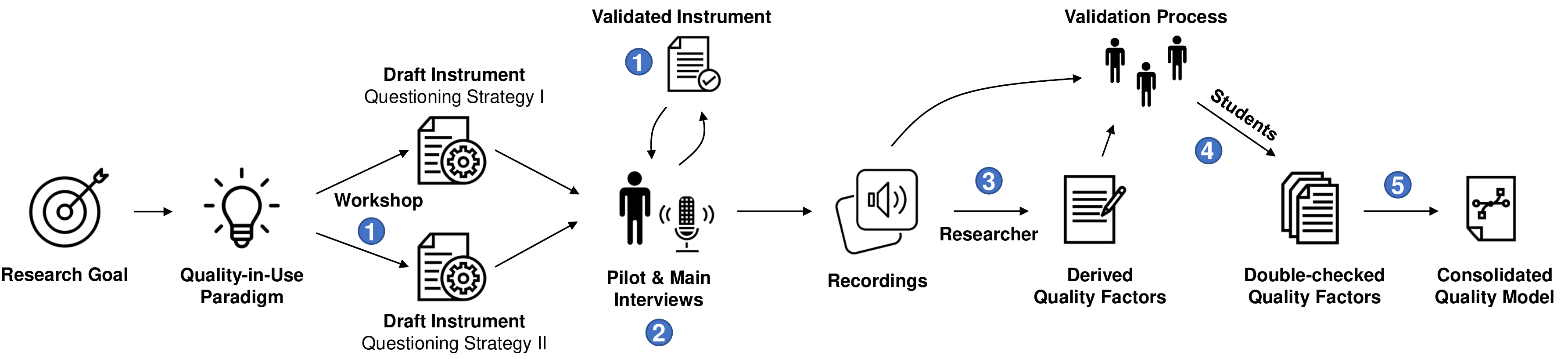}
    \end{mdframed}
    \vspace{-.2cm}
    \caption{Overview of the method followed in our industrial survey: (1) preparing and validating the instrument, (2) conducting interviews, (3) qualitative content analysis, (4) review process, and (5) creating the Activity-Based Artifact Quality Model.}
    \label{fig:approach}
    \vspace{-.4cm}
\end{figure*}

Since most of our interview questions are open-ended, we decided to use qualitative content analysis in order to analyze the interview data (see step \circled{3} in Fig.~\ref{fig:approach}). 
Following the guidelines of Mayring~\cite{Mayring2014}, we conducted a content analysis inductively as our research goal is explorative and we needed to derive the quality factors of the test artifacts from the interview data.
The first author analyzed the interview recordings and performed two steps for each interview. 
First, for each artefact discussed in the interview, the mentioned problems were determined. 
Second, we derived factors from theses problems that influence the quality of the artifacts positively or negatively. 
In particular, we studied the answers to the probing questions, which give a precise insight into which quality factors are considered good or bad.
To validate our results, we performed an internal review process (see step \circled{4} in Fig.~\ref{fig:approach}).
We involved three students and provided them with the interview recordings as well as the hypotheses and derived quality factors.
They performed the same steps as the first author and compared the results in order to agree on the information to be extracted. 
In case of deviations, the respective passages in the interview were analyzed together until reaching a consensus.
After the validation process, we used frequency analysis to find out which problem was mentioned more often. 
This provides a first indication of where systematic quality control may be most needed.

\subsection{Survey Packaging}
We report our results in two ways.
Firstly, in the form of a research paper to share our findings regarding quality control of Agile test artifacts in the research community.
Secondly, as an executive summary to share the results with the interviewed practitioners.

\section{Results}
This section presents the results of our survey structured according to the research questions.
Based on our activity-based quality understanding, we describe for each test artifact (see \textbf{RQ 3}) the stakeholders using it (see \textbf{RQ 1}) and the context of the activities (see \textbf{RQ 2}). 
As described in Section~\ref{questioning}, we applied a ``problem oriented'' questioning approach to determine the quality factors.
We report the identified challenges arising when the stakeholders use each artefact during their activities.
From these challenges, we derived the factors that positively (see \textbf{RQ 4}) or negatively (see \textbf{RQ 5}) influence the quality of the artifacts.
A positive impact of a quality factor is indicated by $\oplus$, while a negative impact is indicated by $\ominus$.
The artifacts that were discussed by our participants were \textsc{acceptance criterion}, \textsc{acceptance test}, \textsc{feature}, \textsc{test documentation}, \textsc{test data}, \textsc{unit test} and, finally, \textsc{all test artifacts} for factors independent of the concrete artifact.

Our final ABAQM (see Fig.~\ref{fig:abaqm}) includes 16 quality factors. Most of the quality factors support the stakeholders in carrying out their activities (13 out of 16 quality factors). However, three quality factors hinder the stakeholders in performing certain activities.

\subsection{Artifact 1: Acceptance Criterion}

\subsubsection{Stakeholder and Activities} \textsc{Acceptance criteria} (AC) are conditions that a system must meet in order to fulfill a \textsc{User story} and be ultimately accepted by the user. AC are used by \textsc{Test designers} during  \textsc{Acceptance test creation}.
This activity involves two steps.
Firstly, the \textsc{Test designer} analyzes all AC assigned to a particular \textsc{User story} to understand the expected system behaviour.
Secondly, the \textsc{Test designer} derives \textsc{Test cases} for each AC and merges them into an \textsc{Acceptance test}, which is later used in the quality assurance process to check the compliance of the system with the \textsc{User story}.
All \textsc{Test designers} interviewed stated that both steps are performed manually. 

\subsubsection{Challenges and Quality Factors} We found two major challenges with the usage of AC during \textsc{Acceptance test creation}.
From these challenges, we derived six quality factors.

\paragraph{\textbf{Challenge 1}: Acceptance criteria are ambiguously formulated.}

The interviewed participants complained about the poor linguistic quality of the AC.
In many cases, it is not clear ``\textit{what exactly the system is supposed to do, which makes it difficult to derive test cases}'' (P3). 
Hence, \textsc{Test designers} need to contact the \textsc{Business analyst} who defined the criterion and clarify its meaning before they can start with the actual test case creation. 
This leads to delays in the test design process.
According to P2 and P3, it would be helpful if the formulation of an AC is checked prior to the test process to ensure that only testable AC are submitted to the \textsc{Test designer}. 
For this purpose, the \textsc{Test designer} should be involved in the formulation of the AC.
However, this would require additional resources which are very limited in practice as stressed by P2 and P17:

\begin{xquote}
 \textit{``We lack the time to discuss every acceptance criterion with each other. We are \textbf{dependent on the formulation skills} of our business analysts.''} (P2)

\textit{``The quality of acceptance criteria varies from project to project. Some analysts specify them precisely, while some do not. However, checking the formulations \textbf{manually is not possible} due to tight time constraints.''} (P17)
\end{xquote}
Hence, a quality assurance check of ACs should be performed automatically to be suitable for practical use.
In this context, the AC should be reviewed with respect to the following quality factors:

 \begin{description}[wide=0\parindent]
            \item[QF 1: Coreferences $\ominus$] 
            A \textsc{user story} usually contains multiple AC specifying the expected system behaviour. 
            In practice, these AC often contain coreferences (i.e., expressions that refer to the same entities). 
            As the number of AC increases, it becomes difficult to resolve these coreferences correctly, which hinders the test design.
            Hence, AC should not contain coreferences to ensure testability.
            \item[QF 2: Vague phrases $\ominus$] 
             The interviews show that AC are usually defined using unrestricted natural language. 
            The use of natural language is intuitive for \textsc{Business analysts}, but bears the risk of vagueness and ambiguity.
            As already stated by Berry and Kamsties, this can lead to ``\textit{diverging expectations and inadequate or undesirably diverging implementations'}'~\cite{Berry2004}.  
            We found that vague phrases often occur in AC and hinder the \textsc{Acceptance test design}.
            
            \begin{xquote}
                 ``\textit{You often see criteria like ``the system should be able to upload the data quickly.'' What exactly is meant by quickly? You do not know then \textbf{what to test}.}'' (P11)
            
                 ``\textit{A typical example you often see: ``if possible, the system should do xy.'' It is unclear \textbf{what possible means}.}'' (P2)
            \end{xquote}
            
            We refrain from listing all vague phrases in the ABAQM since a number of studies~(e.g.\ \cite{Gervasi2019,femmer2017rapid}) have already dealt with this quality factor in requirements. Instead, we want to explicitly point out that this quality issue is also relevant for \textsc{Acceptance criteria}.
            \item[QF 3: References to emails, calls, and documents $\ominus$]
            Instead of fully documenting the desired system functionality, expressions like ``as discussed by phone'' are often found within the AC.
            This leads to a series of problems.
             First, the AC can only be understood by the stakeholders involved in the call and cannot be converted into test cases by another \textsc{Test designer}. 
            Therefore, the testability of the AC is limited due to the undocumented, implicit knowledge.
            Second, information about the system functionality is lost with changes in the project team and it is no longer known which functionality the created test case was initially supposed to test.
             As a result, there is no traceability between the created \textsc{Test case} and the AC.
            In C1, AC also often contain references to other \textsc{Acceptance criteria} or documents (e.g. ``as described in document x''). 
            Due to the high change dynamics in Agile projects, these references become quickly outdated, which results in gaps in the requirement specification.
\end{description}

\paragraph{\textbf{Challenge 2}: Lack of an overview of dependencies between acceptance criteria}

\textsc{Acceptance test design} involves not only the creation of \textsc{Acceptance tests} for new system functionalities but also the adaptation of existing \textsc{Acceptance tests} to changing customer requirements. 
The latter is essential in order to keep requirements and tests aligned. 
For this purpose, \textsc{Test designers} need to understand the relationship between already implemented \textsc{User stories} and new \textsc{User stories}, and adapt the test suite accordingly. 
If the new \textsc{User story} introduces a new functionality, a new \textsc{Acceptance test} must be created and added to the test suite.
If the \textsc{User story} changes an already implemented functionality, the existing \textsc{Acceptance tests} must be adapted.
However, it is increasingly difficult to identify these relationships due to the high number of \textsc{User stories}. 
The interviews showed that for every requested change in the software a new \textsc{User story} is created, rather than the existing \textsc{User story} changed.
This observation coincides with the results of the study by Hotomski et al.~\cite{hotomski}. 
Consequently, the number of \textsc{User stories} and AC is growing steadily, making it more and more difficult to keep track of them: 

\begin{xquote}
``\textit{If someone adds a new user story to the backlog that changes or overwrites another user story, we don't notice it. So we \textbf{don't know which tests we need to adjust}.}'' (P2) 
\end{xquote}
Instead, separate \textsc{Acceptance tests} are created for each \textsc{User story}, resulting in a test suite constantly increasing in scope and complexity.
When the test suite is executed and some tests fail, ``\textit{we don't know if these tests reveal a real bug in the system or if they are checking old functionality and should have been updated}'' (P2).
This leads to additional effort and therefore high testing costs. A similar situation is found in company C2:

\begin{xquote}
``\textit{We do not know whether some acceptance criteria \textbf{overlap or even contradict each other}. Hence, it sometimes happens that we create contradictory test cases.}'' (P3).
\end{xquote}

\begin{description}[wide=0\parindent]
            \item[QF 4: Conflict-free $\oplus$] 
            A \textsc{Test designer} can only maintain a consistent test suite if the underlying AC are not contradictory.
            Consequently, practitioners require a method that automatically compares AC with each other and reveals inconsistencies.
            This will have a positive impact on both \textsc{Acceptance test creation} and \textsc{updating} as it indicates which \textsc{User stories} the \textsc{Test designer} needs to check, as stressed by P2:
            
            \begin{xquote}
            ``\textit{As a test designer I have to understand where and how the functionality is supposed to change and how I need to adapt my tests. If you could somehow \textbf{automatically display overlaps between acceptance criteria}, that would help me a lot.}'' (P2).
            \end{xquote}
            
            In the case of major changes introduced by the new \textsc{User story}, the existing \textsc{Acceptance test} should be archived and a new \textsc{Acceptance test} created.
            Otherwise, the existing \textsc{Acceptance test} should be adapted.
            This helps to avoid false negatives (i.e., invalid failing tests) during test execution.
            According to P2, P10 and P11, the old \textsc{User story} should be eventually assigned the status ``old'' and linked to the new \textsc{User story}.
            This is essential to enable version control of the test assets as described in Section~\ref{allArtifacts}.  
              \item[QF 5: Unique $\oplus$] 
            In order to prevent the creation of unnecessary tests, it is essential that AC do not describe redundant functionalities.
            If two AC describe the same functionality, the \textsc{Business analyst} needs to be informed and both AC need to be merged.
             \item[QF 6: Link to related Acceptance Criteria $\oplus$]
             \textsc{Acceptance tests} sometimes cover more than one AC, sometimes of more than one \textsc{User story}. Therefore, knowing the relations between AC enables to minimize the overall testing effort since related functionalities can be tested simultaneously. 
             
            \begin{xquote}
            ``\textit{We often noticed afterwards that test cases \textbf{could have been bundled together}, e.g. for acceptance criteria that describe the same UI view.}'' (P17) 
    
            ``\textit{We've been trying to optimize our test suite for some time now. But we fail frequently to create combined test cases for related requirements, because we \textbf{do not know which acceptance criteria belong together}.}'' (P4) 
            \end{xquote}
            
            In order to create such joint tests, the \textsc{Test designer} needs to understand the relationships between the AC and consider them during test case creation.
            Accordingly, the quality of AC is considered good if they are linked to related AC.
            An example are two AC that handle the same input parameters as AC 1: ``If input A then function B'' and AC 2: ``If input A and input B then function C.'' Both AC can be checked with a joint \textsc{Acceptance test}.
\end{description}

\subsection{Artifact 2: Acceptance Test}

\subsubsection{Stakeholder and Activities}
\textsc{Acceptance tests} are instruments used to verify the conformity between user expectations and actual system behavior (\textsc{Acceptance testing}).
Each \textsc{Acceptance test} contains a set of \textsc{Test cases}. 
In practice, there are two ways of \textsc{Acceptance testing}.
Internal \textsc{Acceptance testing} (alpha testing), which is performed by members of the organization that developed the software, and external \textsc{Acceptance testing} (beta testing), which is performed by the customer. 
As we were not able to talk to customers during the study, we examine the quality of \textsc{Acceptance tests} from the perspective of an internal \textsc{Product owner} in the context of alpha testing. 

\subsubsection{Challenges and Quality Factors}
Our interviews reveal two challenges with the usage of \textsc{Acceptance tests} during \textsc{Acceptance testing}.
We derived three quality factors from the identified challenges.

\paragraph{\textbf{Challenge 3}: Acceptance tests contain too many or too few test cases}
We found that \textsc{Acceptance tests} are often not systematically created resulting in incomplete or excessive \textsc{Test cases}.

\begin{xquote}
``\textit{We do not follow any particular procedure in the preparation of acceptance tests. Every test designer does this based on his experience. Of course, such a manual process is prone to errors, because you can \textbf{overlook some cases}. In fact, I have also been in the situation where I forgot test cases.}'' (P3)
\end{xquote}
In the case of missing \textsc{Test cases}, system defects are not (or only partially) detected.
As a result, faulty software is ultimately delivered to the customer, leading to errors in production and lower customer satisfaction.
According to an internal analysis conducted by company C1, 83\% of the system defects at company C1 could have been detected by more complete \textsc{Test cases}.
A similar observation was made in company C4: 

\begin{xquote}
``\textit{We often experience that our live system does not completely fulfill all user stories. This could have been avoided by the \textbf{right acceptance tests}}'' (P6).
\end{xquote}
Instead of systematically determining which \textsc{Test cases} are required to cover an AC, they are usually created based on past experience of the \textsc{Test designer}. 
This makes the test case derivation error-prone and increases the risk of missing test cases, as \textsc{Test designers} ``\textit{tend to only test the positive cases and not the negative ones}'' (P1, P9). 
A major challenge is the complexity of the AC: 
\begin{xquote}
``\textit{We often have to implement highly complex business rules that include a range of parameters. It's hard to decide \textbf{which combinations of parameters should be tested}}'' (P1).
\end{xquote}
This increases the risk of missing \textsc{Test cases}. 
However, not only \textsc{Test cases} are missing, but also superfluous \textsc{Test cases} might be created leading to an increase in the testing effort. 
According to P1, many \textsc{Test designers} are lacking the required qualification and, more importantly, the time for a systematic test case derivation. 
Consequently, there is a great demand for an automated test case derivation from AC to maintain the high development speed.

 \begin{description}[wide=0\parindent]
            \item[QF 7: Positive and negative scenarios $\oplus$] 
              \textsc{Acceptance tests} are only suitable for detecting system defects if they are complete~(i.e. covering all positive and negative \textsc{Test cases}). 
              \item[QF 8: Minimal $\oplus$] 
              Achieving QF 7 is crucial to the quality of an \textsc{Acceptance test}, however, it is also necessary to strike a balance between full test coverage and the number required of \textsc{Test cases}. More specifically, an \textsc{Acceptance test} should contain only the minimum number of \textsc{Test cases} needed to fully cover the AC in order to minimize the required testing effort.
\end{description}

    

    

\paragraph{\textbf{Challenge 4}: Lack of automation of acceptance tests}

The interviews revealed that the degree of test automation is still insufficient in practice. At the lower levels of the test pyramid, such as \textsc{Unit tests} and \textsc{Integration tests}, the execution has mostly been automated. However, \textsc{Acceptance tests} are usually performed manually resulting in large testing efforts.

\begin{xquote}
"\textit{Our acceptance tests are \textbf{always carried out manually}. Therefore, the testing process takes a rather long time and we are highly dependent on how the product owner performs the test.}" (P18)
\end{xquote}
This represents a major challenge, especially as the development project proceeds and the system's functionality increases: 

\begin{xquote}
"\textit{Before each new release, we have to run acceptance tests that check already implemented user stories to avoid regression. The number of tests quickly increases during a project and then you ask yourself \textbf{who is going to execute these tests}? We have to run the new acceptance tests as well.}" (P7)
\end{xquote}
 The reason for the low automation of \textsc{Acceptance tests} stems not from limited tool support, but rather from the fact that many companies still neglect to use them: We found that some smaller companies like C6 have already automated the majority of their \textsc{Acceptance tests}. The problem of insufficient automation occurs mainly in large companies like C1 and C2. According to P9, P1 and P6, this might be due to the culture of these companies, who allegedly refuse to implement new automation tools initially and therefore introduce them with a considerable delay.
 
 \begin{description}[wide=0\parindent]
            \item[QF 9: Automated $\oplus$] 
             In order to cope with the high development speed, \textsc{Acceptance testing} needs to be automated, e.g.\ through tools such as Selenium, Cypris and Robot Framework. 
\end{description}

    

\subsection{Artifact 3: Feature}

\subsubsection{Stakeholder and Activities} A \textsc{Feature} is a specific piece of functionality that is desired by the customer. In case of new or changed \textsc{Features} in the current iteration, the \textsc{Test lead} must verify that no regression on already implemented \textsc{Features} is introduced (\textsc{Regression testing}). With a growing system, the scope of \textsc{Regression testing} also increases, so that running an entire regression test suite is time consuming:

\begin{xquote}
"\textit{In our project, some manual regression tests \textbf{take four days}.}" (P4)
\end{xquote}
A similar picture emerges with automated regression tests performed by continuous integration tools: 

\begin{xquote}
"\textit{We run our automated tests via Travis. The \textbf{Travis build} for our entire application \textbf{takes an entire day}.}" (P5)
\end{xquote}
Such long test suite runs pose a major problem, especially considering the short sprint cycles that are often only two weeks. Selecting the right regression tests is therefore essential in order to minimize the testing effort. Specifically, the \textsc{Test lead} needs to run the regression test for the changed \textsc{Feature} and for all dependent \textsc{Features} to identify potential regressions (\textsc{Regression test selection}). For this purpose, knowledge about \textsc{Feature} dependencies is required.

\subsubsection{Challenges and Quality Factors} We found one major challenge related to the usage of \textsc{Features} during \textsc{Regression test selection}. From this challenge, we derived two quality factors. 

\paragraph{\textbf{Challenge 5}: Lack of an overview of dependencies between features}

In practice, there is no overview of the relationships between \textsc{Features}. As a result, there is a negative impact on the \textsc{Regression test selection} as it is not transparent which regression test runs are necessary. Consequently, practitioners need to test on a risk-based basis: 

\begin{xquote}
"\textit{I execute the regression tests of all those features which I know from \textbf{experience} to be related to the changed feature.}" (P5)
\end{xquote}
This is prone to errors resulting in a growing desire of the \textsc{Test leads} for a "functional structure" in their project, which indicates the relationship between the \textsc{Features}. This allows to understand which features might be impacted by a change of a certain feature and need to be tested.

 \begin{description}[wide=0\parindent]
            \item[QF 10: Link to dependent feature $\oplus$] 
             According to P4 and P5, there are too many \textsc{Features} in the projects to manually track the dependencies between them. Furthermore, existing tools such as Jira do not provide the option of illustrating relationships between \textsc{Features} via links. Hence, there is a need for a method that automatically reveals dependent \textsc{Features} in order to establish the "functional structure".
             \item[QF 11: Link to regression test $\oplus$]
             For the selection of regression tests, the \textsc{Test lead} requires a clear traceability between \textsc{Features} and corresponding regression tests. Hence, each \textsc{Feature} must have a link to a corresponding regression test.
\end{description}

    

    
    

\subsection{Artifact 4: Test Documentation}

\subsubsection{Stakeholder and Activities} 
In addition to \textsc{Regression test selection}, the \textsc{Test lead} is also responsible for \textsc{Test reporting} and \textsc{Estimation planning}. As part of \textsc{Test reporting}, the \textsc{Test lead} has to provide an overview of successful and failed tests after each iteration to track the progress of the development team. For this purpose, a comprehensive \textsc{Test documentation} is required.

\subsubsection{Challenges and Quality Factors} We found one major challenge related to the usage of \textsc{Test documentation} during both \textsc{Test reporting} and \textsc{Estimation planning}. From this challenge we derived two quality factors.

\paragraph{\textbf{Challenge 6}: Test results and effort are not properly documented}
The interviews revealed that there is a common problem that test results are not properly documented at all test levels. Especially on the intermediate test levels such as integration testing there is a lack of an overview of the results. Therefore the reporting is mainly done on unit and acceptance level. 

\begin{xquote}
"\textit{I experienced a number of projects that do not separate between the test levels and consider every technical test as a unit test and simply document \textbf{all results at unit level}.}" (P10)
\end{xquote} 
Thus, it is difficult to identify the bugs at the correct test levels and change the software accordingly.
    
    
 \begin{description}[wide=0\parindent]
            \item[QF 12: Contains passed/failed rates at each test level $\oplus$] 
              Each test type needs to be linked to its respective test result. Specifically, the \textsc{Test documentation} needs to contain the corresponding test result for each \textsc{Unit test}, \textsc{Integration test}, \textsc{System test} and \textsc{Acceptance test} to provide a comprehensive overview of all testing levels at any time during the life cycle of the software.
\end{description}
In addition, we found that not only the test results, but also the test effort is not properly documented, which leads to issues in \textsc{Estimation planning}. This activity aims at planning the resources needed to perform the testing in the next iteration: 

\begin{xquote}
"\textit{It is very difficult to estimate the required number of testers when I join a new project as there is \textbf{no documentation of the required test efforts} from previous iterations}". (P4)
\end{xquote} 
Thus, there is no indication how much working days were needed by the testers involved in the project to validate former \textsc{User stories}, as stressed by P4 and P13:

\begin{xquote}
"\textit{We need some \textbf{key performance indicators} especially for Agile testing that are tracked and documented during the test execution. Based on these, I can plan future test activities.}" (P4)

"\textit{For example, it would be great to know \textbf{how many story points} were implemented and tested in past sprints.}" (P13)
\end{xquote}

 \begin{description}[wide=0\parindent]
            \item[QF 13: Contains testing effort per Story Point $\oplus$] 
              A \textsc{Test lead} needs an overview of the number of \textsc{User stories} implemented in past sprints, their story points and the required test effort. Specifically, it is necessary to document how many working days the testers needed per story point to get an overview of their testing capabilities and estimate future testing efforts accordingly. 
\end{description}

    
    

\subsection{Artifact 5: Test Data}

\subsubsection{Stakeholder and Activities}  In addition to suitable \textsc{Test cases}, \textsc{Test data} is also needed to systematically test the behavior of the system. \textsc{Test data} is therefore required at all test levels. Our interviews indicated that there are two main approaches to decide which stakeholder is responsible for which testing level. In small companies, the \textsc{Software engineer} is usually conducting end-to-end tests, i.e. he is responsible for all test levels:

\begin{xquote}
"\textit{We do not have dedicated tester roles. Our software engineers perform the \textbf{entire process} from unit testing to system testing.}" (P9, P6)
\end{xquote}
In large companies, the test levels are allocated to different roles:

\begin{xquote}
"\textit{Unit tests are written and executed by our software engineers, but we have \textbf{different testers} who perform integration tests or other test types.}" (P18)
\end{xquote}
However, the interviews showed that all roles have an interest in high quality \textsc{Test data}.

\subsubsection{Challenges and Quality Factors} We found one challenge related to the usage of \textsc{Test data} and derived one quality factor.

\paragraph{\textbf{Challenge 7}: Lack of test data to properly test the software}

In practice, the generation of \textsc{Test data} is a great challenge, so that "\textit{we often do not have enough test data or the quality of our test data is poor}". (P1). In this context, poor quality denotes the deviations from real production data. The participants complained that their \textsc{Test data} often does not cover all possible boundary cases that might occur in production, so that the system is not tested under all potential conditions. We found this issue in large companies like C1 as well as in small companies like C6. This is mainly caused by the fact that \textsc{Test data} is not systematically derived from production data, but rather that testers use random test values as \textsc{Test data}.

\begin{xquote}
\textit{"I often see the problem in projects that poor test data is used. Poor means that the developer enters \textbf{arbitrary values} in his unit test, but omits constellations that might occur in practice. Obviously, this leads to errors."} (P7)
\end{xquote}
Hence, the testing of all possible boundary values depends on the experience of the testers. 

 \begin{description}[wide=0\parindent]
            \item[QF 14: Boundary values $\oplus$] 
              To ensure that all potential exceptional conditions are covered during testing, the \textsc{Test data} must contain the same boundary cases as the production data.  
    
    \begin{xquote}
    "\textit{We need a method that learns to generate \textbf{appropriate test data} from production data.}" (P1)
    \end{xquote}
    
     In this context, it is crucial to keep the \textsc{Test data} anonymized, especially when dealing with sensitive data. 
\end{description}

    
    
    

\subsection{Artifact 6: Unit Tests}

\subsubsection{Stakeholder and Activities} \textsc{Unit tests} are usually implemented and used by \textsc{Software engineers} to test individual units of the source code or sets of modules.

\subsubsection{Challenges and Quality Factors} We found one challenge related to the usage of \textsc{Unit tests} and derived one quality factor.

\paragraph{\textbf{Challenge 8}: Inadequate Code Coverage of Unit Tests}

The interviews demonstrated that there are not only problems with functional testing (i.e. poor \textsc{Acceptance testing}) but also problems with testing at lower test levels: 

\begin{xquote}
"\textit{There are always bugs in production that could have been detected \textbf{at lower levels}.}" (P1)
\end{xquote}
The core problem mentioned by the participants is that \textsc{Unit tests} are not created following a certain pattern. Rather, it depends on the developer and the reviewer which \textsc{Unit tests} are created. This leads to a strongly fluctuating quality of the \textsc{Unit tests}. Similar to the automation of \textsc{Acceptance tests}, we found differences between small and large companies. The smaller companies were able to give us an overview of the code coverage of their \textsc{Unit tests}, while larger companies were not aware of the quality of their \textsc{Unit tests}.

 \begin{description}[wide=0\parindent]
            \item[QF 15: Code coverage $\oplus$] 
               To control the quality of \textsc{Unit tests}, 
              code coverage metrics should be applied as they allow to determine how much of the developed code is tested. The participants mentioned arbitrary thresholds~(e.g. 80\%) which they considered useful. 
               
    \begin{xquote}
    "\textit{This does not mean that no errors can occur. But it provides a good \textbf{first overview of the quality} of my unit tests.}" (P7)
    \end{xquote} 
    
\end{description}

    
    

\subsection{All Test Artifacts}\label{allArtifacts} 

In the following, we present a challenge that applies to all test artifacts equally. Hence, the quality factor derived from this challenge is relevant for all presented test artifacts.

\paragraph{\textbf{Challenge 9}: Missing version control of all test assets}

\textsc{Configuration management} is an integral activity to monitor and control the status of software during its life cycle. Version control of source code and automated tests is already anchored in today's business practice and is supported by control systems such as git, allowing to track code changes over time. However, version control of all Agile test artifacts is only partially implemented: 

\begin{xquote}
"\textit{We often don't know what the software is capable of doing or has done at a certain point in time and \textbf{what exactly we have tested}.}" (P1)
\end{xquote}
This poses a problem especially in regulatory environments. For example, companies in the insurance industry need to document the functionality of the different software versions and prove which tests have been performed to verify that functionality. Hence, practitioners "\textit{need the historic information of all test assets.}" (P1, P2).

 \begin{description}[wide=0\parindent]
            \item[QF 16: Documented Status $\oplus$] 
              All test artifacts need to be maintained with an appropriate status. This can be illustrated by a \textsc{User story} and its corresponding \textsc{Acceptance test}. After the creation of a \textsc{User story}, its status is set to "New". It will be set to "Committed" by the developer once its implementation has started. After the implementation, the \textsc{User story}'s status is set to "Resolved" and is finally set to "Done" by the \textsc{Product owner} if the \textsc{Acceptance test} was successful. To monitor the status of the software during its life cycle, it is indispensable that artifacts are archived and not discarded. For example, if a new \textsc{User story} overwrites another one, the old \textsc{User story} including the \textsc{Acceptance tests'} status should be set to "old" and the old user story should reference the new \textsc{User story}. This allows to review the tested functionality and the test results for any given build at any given point in time. 
\end{description}

\begin{figure}
     \centering
        \resizebox{8cm}{!}{
        \includegraphics[trim={1.5cm 0.2cm 1.5cm 0.5cm}]{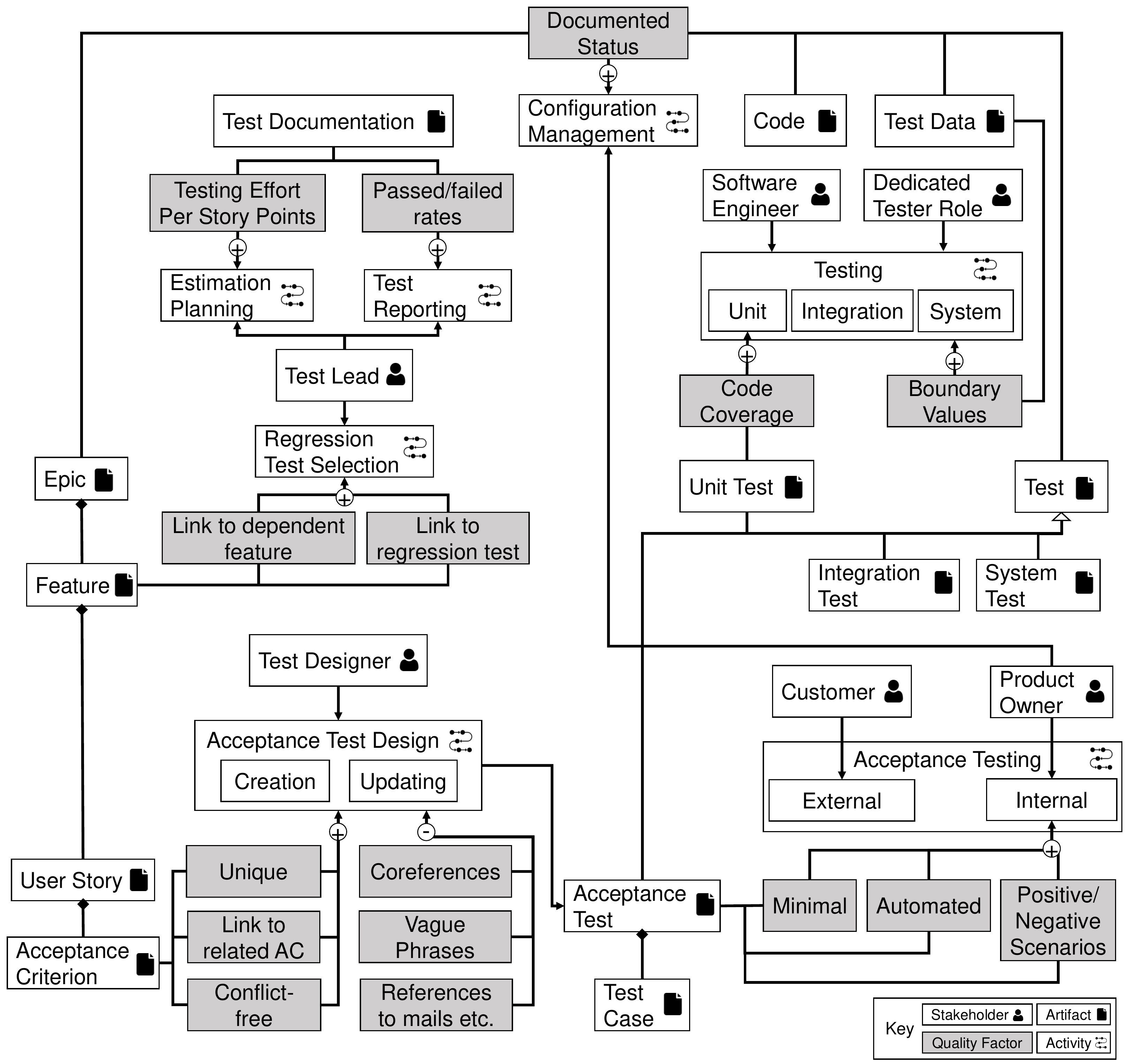}
        }
    \vspace{-.2cm}
    \caption{Activity-Based Quality Model for Agile test artifacts.}
    \label{fig:abaqm}
    \vspace{-.4cm}
\end{figure}

\section{Discussion and Related Work}
In this section, we discuss our findings and put them in context of related work. 

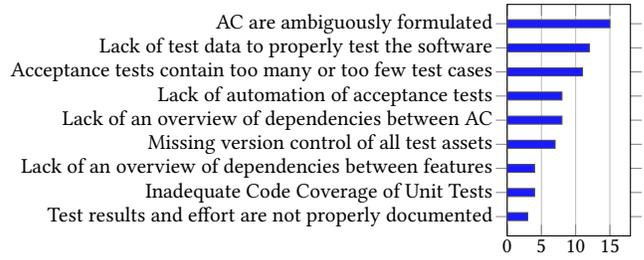
\begin{figure}
\centering
\begin{tikzpicture}
\pgfplotsset{%
    width=.38\columnwidth,
    height=0.55\columnwidth
}

\pgfplotsset{every x tick label/.append style={font=\small, yshift=0.5ex}}
\pgfplotsset{every y tick label/.append style={font=\small, xshift=0.5ex}}

\begin{axis}[ 
    xmajorgrids=true,
    xbar, xmax=18,
    xmin=0,
    symbolic y coords={{Test results and effort are not properly documented}, {Inadequate Code Coverage of Unit Tests}, {Lack of an overview of dependencies between features}, {Missing version control of all test assets}, {Lack of an overview of dependencies between AC}, {Lack of automation of acceptance tests},  {Acceptance tests contain too many or too few test cases},    {Lack of test data to properly test the software}, {AC are ambiguously formulated}},
    ytick=data
]
\addplot[fill=blue!90,draw=black!70,tickwidth = 0pt,bar width=3pt,label style={font=\tiny}, tick label style={font=\tiny}] coordinates {(3,{Test results and effort are not properly documented}) (15,{AC are ambiguously formulated}) (8,{Lack of an overview of dependencies between AC}) (11,{Acceptance tests contain too many or too few test cases}) (8,{Lack of automation of acceptance tests}) (4,{Lack of an overview of dependencies between features})  (12,{Lack of test data to properly test the software}) (4,{Inadequate Code Coverage of Unit Tests}) (7,{Missing version control of all test assets})};
\end{axis}
\end{tikzpicture}
    \vspace{-.5cm}
\caption{Frequency Analysis of Mentioned Challenges.}
\label{fig:frequency}
  \vspace{-.5cm}
\end{figure}

\paragraph{\textbf{Finding 1:} Identified challenges are partly similar to known problems from non-Agile projects}
Poor formulation of requirements and lack of traceability between artifacts are well-known problems from traditional projects~\cite{mendez15}. Interestingly, we have also identified these two problems in our survey. The lack of traceability was observed especially for the artifacts AC and \textsc{Feature}. According to our frequency analysis, the lack of traceability between AC was mentioned by eight of the 18 respondents, whereas the missing traceability between \textsc{Features} was mentioned by only four respondents (see Fig.~\ref{fig:frequency}). The challenge of inadequate formulated \textsc{Acceptance criteria} was mentioned by more than 80\% of the respondents and therefore has the highest frequency in our sample. Our results indicate that already known problems from traditional software projects could not be solved by the shift to Agile software development. Regardless of the development paradigm applied, practitioners seem to face the same quality issues in some of their used test artifacts.

\textbf{Implications for Practice \& Academia}
If teams switch to an Agile paradigm, common requirements and test engineering problems still need explicit attention and won't go away by themselves. Academia will continue evaluating whether solutions applied in traditional software engineering, also work in ASD.


\paragraph{\textbf{Finding 2:} Quality model contains only currently relevant quality factors}
There already are a number of studies on quality factors of artefacts used in traditional projects. For example, it exists a large body of work on quality dimensions of data~\cite{wand96,wang96,bovee03}. An integrated view is provided by Catarci and Scannapieco~\cite{catarci03}, who define the quality of data by the criteria accuracy, completeness, consistency, and timeliness. We were surprised that the interviewed practitioners mentioned only a few of the already known quality factors and instead emphasized specific factors. For example, in the context of \textsc{test data}, the practitioners only mentioned boundary values that the \textsc{Test data} must cover. Completeness of \textsc{Test data}, i.e. "\textit{the degree to which a given data collection includes data describing the corresponding set of real-world objects}"~\cite{catarci03}, seems to be a critical problem in Agile testing. The other quality factors were not discussed. We assume that the practitioners have less pressing problems in maintaining these factors and therefore do not mention them explicitly. Our problem-oriented question approach focuses on current and critical problems. As a result, our quality model contains only those quality factors that are difficult to achieve by the practitioners. This can be seen as a strength, but also as a weakness, since it will produce an incomplete model, yet provide a quality model of what is most relevant. Since project pressure is such a dominant topic in our interviews, we would argue that this makes the model more useful for practitioners.

\textbf{Implications for Practice \& Academia}
If practitioners do not have enough time for full-blown QA of test artifacts, we suggest to start with the quality factors mentioned by fellow practitioners. Academia however, needs to validate the quality model in particular regarding relative relevance of the quality factors. 

\paragraph{\textbf{Finding 3:} Most identified quality factors cannot be controlled manually}
Multiple studies have shown that the high change dynamics and development speed in Agile projects requires an increasing automation of the test process. For example, Fischbach et al. stresses the need for an automatic \textsc{Test case} derivation from \textsc{Acceptance criteria}~\cite{fischbach20}. Our survey indicates that quality control of test artifacts should also be automated as far as possible. We identified a number of quality factors which should be controlled since they have a significant impact on testing activities. However, they cannot be managed manually due to time constraints. The interviewed practitioners are aware of many of the identified quality factors, but cannot meet them without automated tool support. The need for tool support primarily concerns all quality factors related to traceability such as QF 4, 5, 6, and 10.

\textbf{Implications for Practice \& Academia}
Academia and practice need to collaborate on creating effective and efficient tool support for automatic quality control of test artifacts.

\paragraph{\textbf{Finding 4:} The quality of test artifacts influences the quality of other artifacts indirectly}
We reported which quality factors of a test artifact have a positive or negative impact on certain activities. Our interviews revealed, however, that the presence or absence of the identified quality factors not only affects the activity itself but also its output and thus another test artifact which is used in subsequent activities. This can be illustrated by the two artifacts AC and \textsc{Acceptance test}. We identified seven quality factors of a AC, which have an impact on \textsc{Acceptance test design}. The output of this activity are \textsc{Acceptance tests} (see Fig.~\ref{fig:abaqm}). Consequently, the quality of \textsc{Acceptance tests} is indirectly impacted by the quality of AC as they influence the activity in which \textsc{Acceptance tests} are created. This reflects the common claim that quality defects in early artifacts (e.g. requirements-like artifacts) have consequences across multiple layers of indirection.

\textbf{Implications for Practice \& Academia}
Practitioners should carefully analyze which artifacts are at the beginning of the processes and focus their limited QA resources on these artifacts, in particular on AC. Academia should try to understand this in more depth and further qualify and quantify the impact.

\section{Threats to Validity}
\paragraph{Internal Validity}
The interviewees may have misunderstood the questions resulting in poor quality or invalid answers. To minimize this threat, we followed the guidelines by Ciolkowski~\cite{Ciolkowski2003} in the creation of the questionnaire. In addition, we conducted a pilot phase to validate the questionnaire internally through discussions in the research team and externally through pilot interviews. Another threat is that the interviewed practitioners may not have the necessary knowledge to provide suitable input to our study. We minimized this threat by selecting practitioners based on previously defined criteria to ensure sufficient experience. As in every interview-based survey, practitioners' statements may be incorrect due to fear, pride or other subjective biases, despite us stressing the anonymity of the study. As such, our resulting quality model reflects the subjective views on quality and needs to be validated with experiments. The selection bias is another threat to internal validity. Although we have started with personal contacts to find participants, the sampling process has been extended by indirect contacts (snowball sampling). As a result, the selection bias threat has been reduced. Our study is also subject to a potential researcher bias, because all interviews and the data analysis were conducted only by the first author. To minimize this threat, all interviews were audio recorded to document the results of the interviews and to provide a basis for further analysis. In addition, the hypotheses and quality factors derived by the first author were validated by an internal review process in order to mitigate confirmation bias. Furthermore, we assured credibility by sending the identified quality factors to the participants for validation (member checking).
\paragraph{Construct Validity}
The questionnaire might not sufficiently cover our research questions limiting the availability of data that provides suitable answers to the research questions. To minimize this threat, we performed two mitigation actions. First, we designed the questionnaire to successively identify the individual elements of the ABAQM. In addition, we mapped the questions of the questionnaire to the research questions and discussed in the research group if the questions are adequate or if further questions are required to answer the RQ in a targeted way.  
\paragraph{Reliability}
As in every interview-based survey, the limited sample size and the sampling strategy do not provide the statistical basis to generalize the results of the study beyond the studied companies and stakeholders. However, we tried to interview practitioners in different roles from different domains and companies of different sizes to obtain a comprehensive picture of the quality of the test artifacts. Nevertheless, the results of our frequency analysis are not statistically representative and do not allow a general conclusion about challenges in using test artifacts. In order to achieve reasonable generalizabilty, future studies should investigate our derived hypotheses in the context of a broader survey and assess their relevance, e.g. by using a Likert scale. 

\section{Conclusion}
Quality of test artifacts matters. In this paper, we conducted an industrial survey to create an Activity-Based Artifact Quality Model to define what this means from a stakeholder's viewpoint. Specifically, we explored quality factors of test artifacts that have a positive or negative impact on the activities of Agile testers. Our quality model contains 16 quality factors for six test artifacts that are reportedly relevant to at least five stakeholders in the process. Further studies should validate the findings, extend the quality model and research the objective relevance of the mentioned quality factors. We encourage Agile testers to use our quality model as the foundation for systematic quality control in practice. 


\bibliography{references}

\end{document}